\documentclass[showpacs,twocolumn,floats,superscriptaddress]{revtex4}
\usepackage{graphicx}
\usepackage{bm}
\begin{document}
\title{Semiclassical time-evolution of the reduced density
matrix and dynamically assisted generation of entanglement
for bipartite quantum systems}
\author{Ph.~Jacquod}
\affiliation{D\'epartement de Physique Th\'eorique,
Universit\'e de Gen\`eve, CH-1211 Gen\`eve 4, Switzerland}
\date{February 16, 2004}
\begin{abstract}
Two particles, initially in a product state, become entangled when they come
together and start to interact. Using semiclassical methods, we 
calculate the time evolution of the corresponding 
reduced density matrix $\rho_1$, obtained by integrating out the
degrees of freedom of one of the particles. 
To quantify the generation of entanglement, we
calculate the purity ${\cal P}(t)={\rm Tr}[\rho_1(t)^2]$. 
We find that entanglement generation sensitively depends
(i) on the interaction potential, especially on its strength
and range, and (ii) on the nature of the underlying classical dynamics.
Under general statistical assumptions, and for short-ranged 
interaction potentials, we find that ${\cal P}(t)$ decays
exponentially fast if the
two particles are required to interact in a chaotic environment, whereas
it decays only algebraically in a regular system. 
In the chaotic case, the decay rate is given by the golden rule spreading of
one-particle states due to the two-particle coupling, 
but cannot exceed the system's Lyapunov exponent.
\end{abstract}
\pacs{05.45.Mt,03.65.Ud,05.70.Ln,03.67.-a \\[-6mm]}
\maketitle 
{\it When two systems (\ldots) enter into temporary interaction
(\ldots), and when after a time of mutual influence the systems separate 
again, then they can no longer be described in the same way as before,
viz. by endowing each of them with a representative of its own.} 
This is how entanglement was characterized by  
Schr\"odinger almost seventy years ago \cite{Erwin}. 
Entanglement is arguably the most puzzling property       
of multipartite quantum systems, and often leads to
counterintuitive predictions due to, in Einstein's words, 
{\it spooky action at a distance}. Entanglement
has received a renewed, intense interest in recent years in the context 
of quantum information theory \cite{quantcompute}.

In the spirit of Schr\"odinger's above formulation, one is naturally led to
ask the following question: ``What determines the rate of entanglement 
production in a dynamical system ?''. Is this rate 
mostly determined by the interaction
between two, initially unentangled particles, or does it depend on the
underlying classical dynamics ? This is the question we
address in this paper. Previous attempts to answer this question have mostly
focused on numerical investigations \cite{furuya98,miller99,lak01,tanaka02,
Prosen03.1,scott03}, with claims that entanglement is
favored by classical chaos, both in the rate it is 
generated \cite{furuya98,miller99,Prosen03.1} and in the maximal amount it
can reach \cite{lak01}. In particular, strong numerical evidences have been
given by Miller and Sarkar for an entanglement production rate given by the 
system's Lyapunov
exponents \cite{miller99}. These findings have 
however been recently challenged by Tanaka et al. \cite{tanaka02}, 
whose numerical
findings show no increase of the entanglement production rate upon increase
of the Lyapunov exponents in the strongly chaotic but weakly coupled regime,
in agreement with their analytical calculations
relating the rate of entanglement production to classical time correlators.
Ref.~\cite{tanaka02} is seemingly 
in a paradoxical disagreement with the almost identical
analytical approach of Ref.~\cite{Prosen03.1}, where entanglement
production was found to be faster in chaotic systems than in
regular ones \cite{caveat1}. This controversy 
thus calls for a better analytical understanding of the problem.

We present a semiclassical calculation of the time-evolved density
matrix $\rho(t)$ for two interacting, distinguishable particles. In the usual
way, entanglement is quantified by the properties of the reduced
density matrix $\rho_1(t) \equiv {\rm Tr}_2 [\rho(t)]$, obtained 
from the two-particle density matrix by
tracing over the degrees of freedom of one (say, the second) particle.
At $t=0$, the two particles are in a product state of two narrow
wavepackets. We will quantify entanglement by
calculating the purity ${\cal P}(t) \equiv {\rm Tr}[\rho_1(t)^2]$ 
which varies from 0 for fully
entangled to 1 for factorizable 
two-particle states \cite{zurek} (the discussion of how
this could translate into the violation of a Bell 
inequality \cite{bell} is postponed to later works). Compared to other
measures of entanglement such as the von Neumann entropy 
or the concurrence, the purity presents the advantage of being analytically 
tractable. For the weak coupling situation we are interested in 
here, numerical works have moreover shown that von Neumann
and linear entropy $S_{\rm lin}\equiv 1-{\cal P}(t)$  
behave very similarly \cite{tanaka02}.
We thus expect the purity to give a faithful and generic
measure of entanglement.
We note that our semiclassical approach is straightforwardly extended 
to the case of undistinguishable 
particles, provided the nonfactorization of the reduced density  
matrix due to particle statistics
is properly taken care of \cite{yushi}.

Our approach is reminiscent of the semiclassical methods developed
by Jalabert and Pastawski \cite{Jal01,Jac03} (see also
\cite{Tom02}) in the context of the Loschmidt
Echo. We will show how the off-diagonal matrix elements of $\rho_1$ are
related to classical action correlators, in a similar way as in Ref.~
\cite{tanaka02,Prosen03.1}. 
Under nonrestrictive statistical assumptions, 
we find that, following an initial transient where $\rho_1$
relaxes but remains almost exactly pure, entanglement
production is exponential in chaotic systems, while it is algebraic
in regular systems. The asymptotic rate of entanglement
production in chaotic systems depends on the strength of the interaction
between the two particles, and is explicitely given by a
classical time-correlator. We note that, 
as is the case for the Loschmidt Echo \cite{Jac01,Cerruti03},
this regime is also adequately captured by an approach based on 
Random Matrix Theory (RMT) \cite{Jac04} -- the time-correlator is 
then replaced by the
golden rule spreading of one-particle states due to the interaction. 
For stronger coupling however,
the dominant stationary phase solution becomes interaction independent,
and is determined only by the classical dynamics, the system's Lyapunov 
exponents giving an upper bound for the rate
of entanglement production. 
The crossover between the two regimes
occurs once the golden rule width becomes comparable to the system's Lyapunov
exponent. Still, one has to keep in mind
that long-ranged interaction potentials can lead to 
significant modifications of this picture, especially at short
times, due to an anomalously slow 
vanishing
of off-diagonal matrix elements of $\rho_1$ within a bandwidth set
by the interaction range.

We start with an initial two-particle product state
$|\psi_1 \rangle \otimes |\psi_2 \rangle \equiv
|\psi_1, \psi_2 \rangle$. The state of each particle is a
Gaussian wavepacket $\psi_{1,2}({\bf y}) = 
(\pi \sigma^2)^{-d/4} \exp[i {\bf p}_0 \cdot ({\bf y}-{\bf r}_{1,2})-
|{\bf y}-{\bf r}_{1,2}|^2/2 \sigma^2]$. 
We write the two-particle Hamiltonian as $
{\cal H} = H \otimes I + I \otimes H + {\cal U}$,
i.e. the two particles are subjected to the same Hamiltonian dynamics. 
At this point, we only
specify that
the interaction potential ${\cal U}$ is smooth, depends only on the
distance between the particles, and is characterized by 
a typical length scale $\zeta$ (which can be its range, or the scale
over which it fluctuates). Setting $\hbar \equiv 1$,
the two-particle density matrix evolves according to
$\rho(t) = \exp[-i {\cal H} t] \rho_0 \exp[i {\cal H} t]$
starting initially with $\rho_0=|\psi_1,\psi_2\rangle\langle \psi_1,\psi_2|$. 
The elements 
$\rho_1({\bf x},{\bf y};t) = \int d{\bf r} \langle {\bf x}, {\bf r}|
\rho(t) |{\bf y}, {\bf r} \rangle$ of the reduced density matrix read 
\begin{widetext}
\begin{eqnarray}
\rho_1({\bf x},{\bf y};t) &=& (\pi \sigma^2)^{-d}
\int d{\bf r} \int \Pi_{i=1}^4 d{\bf y}_i 
\exp[-\{({\bf y}_1-{\bf r}_1)^2 
+({\bf y}_2-{\bf r}_2)^2 + ({\bf y}_3-{\bf r}_1)^2 + ({\bf y}_4-{\bf r}_2)^2 \}
/2 \sigma^2] \nonumber \\
&& \times \;\exp[i {\bf p}_0({\bf y}_1+{\bf y}_2-{\bf y}_3-{\bf y}_4)] \;
\langle {\bf x}, {\bf r}| \exp[-i{\cal H}t] |{\bf y}_1, {\bf y}_2
\rangle
\langle {\bf y}_3, {\bf y}_4 |
\exp[i {\cal H}t] |{\bf y}, {\bf r} \rangle.
\end{eqnarray}
We next introduce the semiclassical two-particle propagator
\begin{eqnarray}
 \langle {\bf x}, {\bf r}| \exp[-i{\cal H}t] |{\bf y}_1,
{\bf y}_2 \rangle &=& 
(-i)^d \; \sum_{s,s'} C_{s,s'}^{1/2} \exp[i \{S_s({\bf y}_1,{\bf x};t) +
S_{s'}({\bf y}_2,{\bf r};t) +
{\cal S}_{s,s'}({\bf y}_1,{\bf x};{\bf y}_2,{\bf r};t)
-\frac{\pi}{2} (\mu_{s}+\mu_{s'}) \}], \nonumber 
\end{eqnarray}
which is expressed as a sum over pairs of classical trajectories, labelled $s$
and $s'$, respectively 
connecting ${\bf y}_1$ to ${\bf x}$ and ${\bf y}_2$ to ${\bf r}$
in the time $t$. Each such pair of paths gives a 
contribution containing one- (denoted by $S_s$ and $S_{s'}$) 
and two-particle (denoted by ${\cal S}_{s,s'}=\int_0^t dt_1 {\cal U}({\bf q}_s
(t_1), {\bf q}_{s'}(t_1))$ \cite{caveat})
action integrals
accumulated along $s$ and $s'$, a pair of Maslov
indices $\mu_{s}$ and $\mu_{s'}$, 
and the determinant $C_{s,s'}$ of the stability
matrix corresponding to the two-particle 
dynamics in the $(2d)-$dimensional space. 
(With the above definition, $C_{s,s'}$ is real and positive.)
We consider sufficiently
smooth interaction potentials varying over a distance much 
larger than $\sigma$. We thus set ${\cal S}_{s,s'}({\bf y}_1,{\bf x};
{\bf y}_2, {\bf r};t) \simeq {\cal S}_{s,s'}({\bf r}_1,{\bf x};{\bf r}_2, 
{\bf r};t)$ (still we keep in mind that ${\bf r}_{1}$ and
${\bf r}_{2}$, taken as arguments of
the two-particle action integrals, have an uncertainty $O(\sigma)$), 
and use the narrowness of the initial wavepackets to linearize
the one-particle actions in ${\bf y}_i-{\bf r}_j$ ($i=1,\ldots 4$; $j=1,2$).
We consider the weak coupling regime, where the one-particle actions
vary faster than their two-particle counterpart. We thus perform a 
stationary phase approximation on the $S$'s to get the
semiclassical reduced density matrix as 
\begin{eqnarray}
\label{semicrho}
\rho_1({\bf x},{\bf y};t) &=& (-4 \pi \sigma^2)^d \sum_{s,l} \;
\exp[i\{S_s({\bf r}_1,{\bf x};t)-S_{l}({\bf r}_1,{\bf y};t)\}] \nonumber \\
&& \times \; \int d{\bf r}
\sum_{s'} {\cal M}_{s,s'} {\cal M}_{l,s'}^{\dagger}
\; \exp[i\{{\cal S}_{s,s'}({\bf r}_1,{\bf x};{\bf r}_2,{\bf r};t)
-{\cal S}_{l,s'}({\bf r}_1,{\bf y};{\bf r}_2,{\bf r};t)\}], \\
{\cal M}_{s,s'} &=& C_{s,s'}^{1/2} \; \exp[-i \frac{\pi}{2}(\mu_{s}-\mu_{s'})]
\; \exp[-\frac{\sigma^2}{2} \{
({\bf p}_{s}-{\bf p}_0)^2 + ({\bf p}_{s'}-{\bf p}_0)^2\}]. \nonumber 
\end{eqnarray}
It is straightforward to see that 
${\rm Tr}[\rho_1(t)]=1$ and $\rho_1({\bf x},{\bf y};t)=[\rho_1({\bf y},
{\bf x};t)]^*$, as required. 
Enforcing a further stationary phase condition on Eq.~(\ref{semicrho})
amounts to performing an average 
over different initial conditions ${\bf r}_{1,2}$. 
It results in $s=s'$, ${\bf x}={\bf y}$, and thus
$\langle \rho_1 ({\bf x},{\bf y};t) \rangle = \delta_{{\bf x},{\bf y}}/\Omega$ 
($\Omega$ is the system's volume), i.e. only
diagonal elements of the reduced density matrix have a nonvanishing average
(the ergodicity of $\langle \rho_1 ({\bf x},{\bf x};t) \rangle$
is due to the average over initial
conditions). For each initial condition, $\rho_1$
has however nonvanishing off-diagonal matrix elements, 
with a zero-centered distribution whose
variance is given by $\langle 
\rho_1({\bf x},{\bf y};t)  \rho_1({\bf y},{\bf x};t) \rangle$.
Squaring Eq.~(\ref{semicrho}), averaging over ${\bf r}_{1,2}$ and
enforcing a stationary phase approximation on the $S$'s, one gets
\begin{eqnarray}\label{Sigma2}
&&\langle 
\rho_1({\bf x},{\bf y};t)  \rho_1({\bf y},{\bf x};t) \rangle=
(4 \pi \sigma^2)^{2d}  \int d{\bf r} d{\bf r}'
\sum_{s,s'} \; \sum_{l,m} \;
{\cal M}_{s,s'} \; {\cal M}_{l,m} \;
{\cal M}_{l,s'}^{\dagger} \; {\cal M}_{s,m}^{\dagger} \;
\langle {\cal F} \rangle, \\
\label{Faction}
&&{\cal F} =
\exp[i \{ {\cal S}_{s,s'}({\bf r}_1,{\bf x};{\bf r}_2,{\bf r};t) 
- {\cal S}_{l,s'}({\bf r}_1,{\bf y};{\bf r}_2,{\bf r};t) \}] \;
\exp[i \{ 
{\cal S}_{l,m}({\bf r}_1,{\bf y};{\bf r}_2,{\bf r}';t)
- {\cal S}_{s,m}({\bf r}_1,{\bf x};{\bf r}_2,{\bf r}';t) \}].
\end{eqnarray}
Our analysis of Eqs.~(\ref{Sigma2}-\ref{Faction}) starts by noting that
$\langle |\rho_1|^2\rangle$ is given by the sum of two positive
contributions.
First, those particular paths for which ${\bf r}={\bf r}'$ and $s'=m$, 
accumulate no phase (${\cal F}=1$) and thus have to be considered
separately. On average, their contribution does not depend on 
${\bf x}$ nor ${\bf y}$, and decays in time only because of their decreasing 
measure with respect to all the paths with ${\bf r} \ne {\bf r}'$. 
Their average contribution is given by 
\begin{eqnarray}\label{bounddecay}
\Sigma^2_0(t) &=& (4 \pi \sigma^2)^{2d} \; \Omega^{-2}
\int d{\bf r}_1 d{\bf r}_2 d{\bf r} \sum_{s,l,m} \;
|{\cal M}_{s,m}|^2 \; |{\cal M}_{l,m}|^2 
\propto \left\{ 
\begin{array}{cc}
\Omega^{-2} \exp[-(\lambda_1+\lambda_2) t]  \;\;\; ; \;\;\; {\rm chaotic}, \\
\Omega^{-2} \left(\frac{t_0}{t}\right)^2 
\;\;\;\;\;\;\;\;\;\;\;\;\;\;\;\;\;\;\;\; ; \;\;\; {\rm regular}.
\end{array}
\right.
\end{eqnarray}
This is obtained by taking half of the $C$'s in the 
${\cal M}$'s as Jacobians for the coordinate transformation 
from ${\bf r}$'s to ${\bf p}$'s, and replacing the other half by
the large time limit \cite{Ozorio}
$C_{s,m} \equiv C_s^{(1)} C_m^{(2)} \propto (t_0/t)^2$ for regular,
and $\approx \exp[-(\lambda_1+\lambda_2) t]$, for chaotic
systems (In Eq.~(\ref{bounddecay}), we explicitely wrote
particle indices; in our
case where the two particles are subjected to the same Hamiltonian,
the Lyapunov exponents are equal, $\lambda_1=\lambda_2$).
Two facts are noticeable here: these contributions do not
depend on the interaction strength, moreover, Eq.~(\ref{bounddecay}) 
gives a lower bound for the decay of $\langle |\rho_1|^2 \rangle$.

The second, generic contributions to $\langle |\rho_1|^2 \rangle$ 
decay in time with $\langle {\cal F} \rangle$.
From Eq.~(\ref{Faction}), it is natural to expect that 
$\langle {\cal F} \rangle$ 
is a decreasing function of $|{\bf x}-{\bf y}|$ and $t$ only. 
Sums and integrals in Eq.~(\ref{Sigma2}) can then be
performed separately to get $\langle 
\rho_1({\bf x},{\bf y};t)  \rho_1({\bf y},{\bf x};t) \rangle
= \Sigma^2_0(t) + \langle {\cal F}({\bf x},{\bf y};t) \rangle/\Omega^2$, with
\begin{eqnarray}\label{2correl}
\langle {\cal F} ({\bf x},{\bf y};t) \rangle &=&\exp\left[
-2 \int_0^t dt_1 dt_2  \; \langle 
{\cal U}({\bf q}_s(t_1), {\bf q}_{s'}(t_1))\;
{\cal U}({\bf q}_s(t_2);{\bf q}_{s'}(t_2)) \rangle \right] \times \nonumber \\
&&\exp\left[ 
+2 \int_0^t dt_1 dt_2 \; \langle 
{\cal U}({\bf q}_s(t_1), {\bf q}_{s'}(t_1))\;
{\cal U}({\bf q}_l(t_2);{\bf q}_{s'}(t_2)) \rangle \right] .
\end{eqnarray}
The behavior of $\langle {\cal F}({\bf x},{\bf y};t) \rangle$ sensitively
depends on the distance $|{\bf x}-{\bf y}|$ between the endpoints of
the classical trajectories ${\bf q}_{s}(t)$ and ${\bf q}_{l}(t)$
connecting ${\bf r}_1$ to ${\bf x}$ and ${\bf y}$ respectively. 
For small $|{\bf x}-{\bf y}|$, the second line on the right-hand side of
Eq.~(\ref{2correl}) almost compensates the first one \cite{Cucchietti}, and 
a Taylor expansion of the difference of the 
two-particle action integrals in Eq.(\ref{Faction}) gives in lowest
order
\begin{eqnarray}\label{Fsmall}
\langle {\cal F} (|{\bf x}-{\bf y}| \le \zeta;t)\rangle
&=& \exp\left[-2 \sum_{\alpha,\beta=1}^d
({\bf x}-{\bf y})_{\alpha} ({\bf x}-{\bf y})_{\beta}
\int_0^t dt_1 \; dt_2 \langle \partial_{\alpha}^{(s)} {\cal U}
({\bf q}_s(t_1), {\bf q}_{s'}(t_1)) \;
\partial_{\beta}^{(s)} {\cal U}({\bf q}_s(t_2);{\bf q}_{s'}(t_2))
\rangle \right].
\end{eqnarray}
The reduced density matrix has a Gaussian decay away
from the diagonal, which we expect to 
hold for $|{\bf x}-{\bf y}| \ll \zeta$ (the distance over which
the two terms in Eq.~(\ref{2correl}) almost cancel each other), 
and for short enough times $t < \tau$ (after which very different,
uncorrelated classical paths $l$ and $s$ may still reach two
neighboring endpoints).
An estimate for $\tau$ is given by the time it takes for two initial
conditions within a distance $\sigma$ 
(this is the uncertainty in ${\bf r}_1$) 
to move away a distance $\propto \zeta$ from each other.
In a chaotic system, this is the Ehrenfest time 
$\tau = \lambda_1^{-1} \ln(\zeta/\sigma)$, while in a regular
system, $\tau$ is much longer, $\tau = O([\zeta/\sigma]^\beta)$ \cite{Berman}.
At larger distances, $|{\bf x}-{\bf y}| \gg \zeta$,  
$\langle {\cal F} \rangle$ is given by the first term
on the right-hand side of Eq.~(\ref{2correl}). Note that,
because the four classical paths in that term come in two pairs,
the dependence on $|{\bf x}-{\bf y}|$ vanishes.

This concludes our semiclassical calculation of the reduced density
matrix for interacting two-particle systems.
We have learned that the variance of
off-diagonal matrix elements 
of $\rho_1$ is determined by classical correlators, with the
important caveat that they are bound downward by
the expressions given in Eq.~(\ref{bounddecay}). 
For the rest of the discussion, we note that,
provided the correlators in Eqs.~(\ref{2correl}) and (\ref{Fsmall})
decay faster than $\propto |t_1-t_2|^{-1}$, 
the off-diagonal matrix elements exhibit a dominant
exponential decay in time. This condition is rather nonrestrictive
and is surely satisfied in a 
chaotic system. We therefore assume from now on a 
fast decay of the correlations,
\begin{eqnarray}\label{correl1}
\langle 
{\cal U}({\bf q}_s(t_1), {\bf q}_{s'}(t_1))\;
{\cal U}({\bf q}_s(t_2);{\bf q}_{s'}(t_2)) \rangle & = &\Gamma \;
\delta(t_1-t_2), \\
\label{correl2}
\langle \partial_{\alpha}^{(s)} {\cal U}
({\bf q}_s(t_1), {\bf q}_{s'}(t_1)) \;
\partial_{\beta}^{(s)} {\cal U}({\bf q}_s(t_2);{\bf q}_{s'}(t_2))\rangle
&=& \gamma \; \delta_{\alpha,\beta} \; \delta(t_1-t_2).
\end{eqnarray}
\end{widetext}
$ $\\[-7mm]
Entanglement is quantified by the purity ${\cal P}(t)=\int d{\bf x} d{\bf y}
\langle \rho_1({\bf x},{\bf y};t)  \rho_1({\bf y},{\bf x};t) \rangle$, 
which is straightforward to compute
from Eqs.~(\ref{Sigma2}-\ref{correl2}). We get three
distinct regimes of decay: (i) an initial regime of classical
relaxation for $t<\tau$. During that time, $\rho_1$ evolves from
a pure, but localized
$\rho_1(0)=|{\bf r}_1 \rangle \langle {\bf r}_1|$
to a less localized, but still almost pure $\rho_1(t)$, with an
algebraic purity
decay obtained from Eqs.~(\ref{Fsmall}) and (\ref{correl2})
as ${\cal P}(t < \tau) \simeq
\Omega^{-1} (\pi/2 \gamma t)^{d/2}$ (even in the case
of a correlator (\ref{correl2}) saturating at a finite value
for $|t_1-t_2| \rightarrow \infty$, which may occur
in regular systems, this initial decay
will still be algebraic $\propto t^{-d}$); 
(ii) a regime where quantum
coherence develops between the two particles so that
$\rho_1$ becomes a mixture. From Eq.~(\ref{bounddecay}) 
and the first line of Eq.~(\ref{2correl}) with Eq.~(\ref{correl1})
one gets ${\cal P}(t) \simeq \exp[-{\rm min}(2 \Gamma;\lambda_1+\lambda_2) t]$
in a chaotic system; in a regular system one has
${\cal P}(t) \propto (t_0/t)^{-2}$, since
$\Sigma_0^2(t)$ dominates the decay of ${\cal P}(t)$ independently
of the correlators in Eqs.(\ref{correl1}-\ref{correl2}); 
and (iii) a saturation regime where the purity
reaches its minimal value. In the chaotic limit, 
this saturation value can be estimated using 
a RMT approach as ${\cal P}(\infty) = 2(\sigma^d/\Omega)+
O(\Omega^{-2})$ \cite{Jac04}, which is in qualitative agreement with
the results obtained in Ref.\cite{lak01} for the von Neumann
entropy. There is no reason to expect a
universal saturation value in the regular regime. 

Analyzing these results, we note that
Eqs.~(\ref{2correl}) and (\ref{Fsmall}) are reminiscent of
the results obtained for ${\cal P}(t)$ by
perturbative treatments in Refs.~\cite{tanaka02,Prosen03.1}, but they
apply well beyond the linear response regime. Our weak coupling
condition
that the one-particle actions $S$ vary faster than the two-particle
actions ${\cal S}$ roughly gives an upper bound $U \le E$ for the
typical interaction strength $U$ ($E$ is the one-particle energy). 
The linear response regime
is however restricted by a much more stringent condition $U \le \Delta \ll E$
($\Delta$ is the mean level spacing) \cite{Jac01}.
The decay regime
(ii) of ${\cal P}(t)$ reconciles the a priori contradicting claims
of Refs.~\cite{furuya98,miller99,Prosen03.1} 
and Ref.~\cite{tanaka02}. For weak coupling,
the decay of ${\cal P}(t)$ is given by classical correlators, and thus
depends on the interaction strength, in agreement with Ref.~\cite{tanaka02}. 
However, ${\cal P}(t)$ cannot decay faster than the bound
given in Eq.~(\ref{bounddecay}), so that at stronger coupling, and in
the chaotic regime, one recovers the results of Ref.~\cite{miller99}.
Simultaneously, regime (ii) also explains the data in Fig.~2 and 4 of
Ref.~\cite{Prosen03.1}, showing an
exponential decay of ${\cal P}(t)$ in the chaotic regime, and
a power-law decay with an exponent close to 2
in the regular regime (this power-law decay was left unexplained by 
the authors of Ref.~\cite{Prosen03.1}).
Our semiclassical treatment thus presents a unified picture
of the problem.

Three more remarks are in order here. First the stationary phase solutions
leading to the above results still hold in the case when
the two particles are subjected to different one-particle Hamiltonians.
Second, the power-law decay
of ${\cal P}(t)$ predicted above for regular systems,
is to be taken as an average over initial conditions ${\bf r}_{1,2}$
(in that respect see Refs.~\cite{Jac03} and \cite{Prosen03.2}), but
may also hold for individual initial conditions, as e.g.
in \cite{Prosen03.1}. Finally, there are cases when the
correlators (\ref{correl1}) and (\ref{correl2}) decay exponentially in time 
with a rate related to the spectrum of Lyapunov exponents.
This also may induce a dependence of ${\cal P}(t)$
on the Lyapunov exponents, which can be captured by the linear response
approach of Ref.~\cite{tanaka02}. We note however that this is not a
generic situation, as most fully chaotic, but nonhyperbolic
systems have power-law decaying correlations. 

As a concluding line, and
noting similarities between the problem treated here and
that of decoherence by an environment \cite{zurek}, we anticipate that
a semiclassical approach as the one presented here
(see also Ref.~\cite{Heller})
could clarify how decoherence relates to the intrinsic
dynamics of the system \cite{zurek,zurek2}.

This work has been supported by the Swiss National Science Fondation.
We thank Atushi Tanaka for a clarifying discussion of 
Refs.~\cite{miller99,tanaka02}.


\begin{thebibliography}{99}
\bibitem{Erwin} E. Schr\"odinger, Proc. Cambridge Phil. Soc. {\bf 31}, 
555 (1936).
\bibitem{quantcompute} M.A. Nielsen and I.L. Chuang, {\it Quantum Computation 
and Quantum Information}, Cambridge University Press, Cambridge (2000).
\bibitem{furuya98} K. Furuya, M.C. Nemes, and G.Q. Pellegrino, Phys. Rev. 
Lett. {\bf 80}, 5524 (1998).
\bibitem{miller99} P.A. Miller and S. Sarkar, Phys. Rev. E {\bf 60}, 1542 
(1999).
\bibitem{lak01} A. Lakshminarayan, Phys. Rev. E {\bf 64}, 036207 (2001);
J.N. Bandyopadhyay and A. Lakshminarayan, quant-ph/0307134.
\bibitem{tanaka02} A. Tanaka, H. Fujisaki, and T. Miyadera, Phys. Rev. E 
{\bf 66}, 045201(R) (2002); H. Fujisaki, T. Miyadera, and A. Tanaka, 
Phys. Rev. E {\bf 67}, 066201 (2003).
\bibitem{Prosen03.1} M. \v Znidari\v c and T. Prosen, J. Phys. A {\bf 36}, 
2463 (2003).
\bibitem{scott03} A.J. Scott and C.M. Caves, J. Phys. A {\bf 36},
9553 (2003).
\bibitem{caveat1} Care should however be taken when comparing fully
chaotic systems with regular ones, as there is no universality to be
expected in the latter case.
\bibitem{zurek} W.H. Zurek, Rev. Mod. Phys. {\bf 75}, 715 (2003).
\bibitem{bell} J.S. Bell, {\it Speakable and Unspeakable in Quantum 
Mechanics}, Cambridge University Press, Cambridge (1987).
\bibitem{yushi} Yu Shi, Phys. Rev. A {\bf 67}, 024301 (2003).
\bibitem{Jal01} R.A. Jalabert and H.M. Pastawski, Phys. Rev. Lett. 
{\bf 86}, 2490 (2001).
\bibitem{Jac03} Ph. Jacquod, \.I. Adagideli, and C.W.J. Beenakker, 
Europhys. Lett. {\bf 61}, 729 (2003).
\bibitem{Tom02} N.R. Cerruti and S. 
Tomsovic, Phys. Rev. Lett. {\bf 88}, 054103 (2002);
J. Vanicek and E.J. Heller, Phys. Rev. E {\bf 67}, 016211 (2003).
\bibitem{Jac01} Ph. Jacquod, P.G. Silvestrov, and C.W.J. Beenakker,
Phys. Rev. E {\bf 64}, 055203 (R) (2001).
\bibitem{Cerruti03}
N.R. Cerruti and S. Tomsovic, J. Math. Phys. {\bf 36}, 3451 (2003).
\bibitem{Ozorio} A.M. Ozorio de Almeida, {\it Hamiltonian Systems:
Chaos and Quantization}, Cambridge (1988). 
\bibitem{Berman} G.P. Berman and G.M. Zaslavsky, Physica {A \bf 91},
450 (1978); M.V. Berry and N.L. Balasz, J. Phys. A {\bf 12}, 625 (1979).
\bibitem{Cucchietti} Because of the second line in Eq.(\ref{2correl}),
the connection proposed in Ref.~\cite{cucch}
between decoherence and Loschmidt Echo breaks down at short times.
\bibitem{cucch} F.M. Cucchietti, D.A.R. Dalvit, J.P. Paz, and W.H. Zurek,  
Phys. Rev. Lett. {\bf 91}, 210403 (2003).
\bibitem{Jac04} Ph. Jacquod, unpublished (2003).
\bibitem{caveat} If ${\cal U}$ 
factorizes as ${\cal U}=V \otimes I + I \otimes W$, 
${\cal S}_{s,s'}({\bf y}_1,{\bf x};{\bf y}_2,{\bf r};t)=
S^{(U)}_s({\bf y}_1,{\bf x};t) + S^{(W)}_{s'}({\bf y}_2,{\bf r};t)$,
the two-particle action thus vanishes
and no entanglement is generated, as should be.
\bibitem{Prosen03.2} T. Prosen and M. \v Znidari\v c, New Jour. Phys. {\bf 5},
109 (2003).
\bibitem{Heller} G.A. Fiete and E.J. Heller, Phys. Rev. A {\bf 68}, 022112 
(2003).
\bibitem{zurek2} W.H. Zurek, Physics Today {\bf 44}, 36 (1991); 
A.K. Pattanayak, Phys. Rev. Lett. {\bf 83},
4526 (1999).
\end{thebibliography}
\end{document}